\documentclass[11pt,twoside]{article}
\usepackage{graphicx,epsfig,natbib,epstopdf}
\usepackage{CS18}
\begin{document}
%
%
%
\title{Star-planet interactions}
%
%
\author{A.~F.~Lanza}
\affil{INAF-Osservatorio Astrofisico di Catania, \\ 
Via S.~Sofia,~78 -- 95123 Catania, Italy}
%
\begin{abstract}
%
%
Stars interact with their planets through gravitation, radiation, and magnetic fields. I shall focus on the interactions between late-type stars with an outer convection zone and close-in planets, i.e., with an orbital semimajor axis smaller than $\approx 0.15$~AU.  I shall review the roles of tides and magnetic fields considering some key observations and discussing theoretical scenarios for their interpretation with an emphasis on open questions. 
\end{abstract}
%
%
%
%
%
\section{Introduction}

Stars interact with their planets through gravitation, radiation, and magnetic fields. I shall focus on  the case of main-sequence late-type stars and close-in planets (orbit semimajor axis $a \la 0.15$~AU) and limit myself to a few examples. Therefore, I apologize for missing important topics in this field some of which have been covered in the contributions by Jardine, Holtzwarth, Grissmeier, Jeffers, and others at this Conference.  

The space telescopes CoRoT \citep{Auvergneetal09} and Kepler \citep{Boruckietal10} have opened a new era in the detection of extrasolar planets and the study of their interactions with their host stars because they allow us, among others, to measure stellar rotation in late-type stars through the light modulation induced by photospheric brightness inhomogeneities  \citep[e.g., ][]{Afferetal12,McQuillanetal14,Lanzaetal14}. Moreover, the same flux modulation can be used to derive information on the active longitudes where starspots preferentially form and evolve \citep[e.g., ][]{Lanzaetal09a,BonomoLanza12}\footnote{Detailed tests of the capability of spot modelling to recover the actual distributions of the active regions have been performed, e.g.,  for the Sun as a star \citep{Lanzaetal07},  CoRoT-2, using starspot occultations during transits \citep{SilvaValioLanza11}, and  HR~1099, using Doppler imaging \citep{Lanzaetal06}.}. Combining all the information together, we can study the correlation between stellar rotation, spot activity, and close-in planets as well as investigate the relationships between different stellar and orbital parameters as derived  by  follow-up observations. A recent result obtained by studying the rotation of Kepler planetary candidates is the paucity of close-in planets around fast-rotating stars. I shall discuss models to account for this intriguing phenomenon in Sect.~\ref{tides}

Optical spectroscopy allows us to study the emission of stellar chromospheres in lines such as Ca II H\&K or H$\alpha$. Some indication of chromospheric features associated to close-in planets have been reported. Stellar coronae can be monitored through X-ray observations and also in that case there have been reports concerning possible planetary effects. Therefore, I shall consider interactions potentially leading to  chromospheric and coronal features in Sect.~\ref{SPMI} The high-energy radiation emitted by stellar coronae plays a fundamental role in the evaporation of the atmospheres of close-in planets, but other kinds of interactions can play a role as well (cf. Sect.~\ref{evaporation_sec}). The presence of close-in planets may affect the evolution of the stellar angular momentum and this aspect will be briefly discussed in Sect.~\ref{starrotation} Finally, circumstantial evidence for photospheric starspot activity related to close-in planets will be introduced in Sect.~\ref{starspots}, while some general conclusions will be reported in Sect.~\ref{conclusions} 

\section{Some tidal effects  in star-planet systems}
\label{tides}

\begin{figure}[ht!]
\centering
\includegraphics[width=90mm,angle=90]{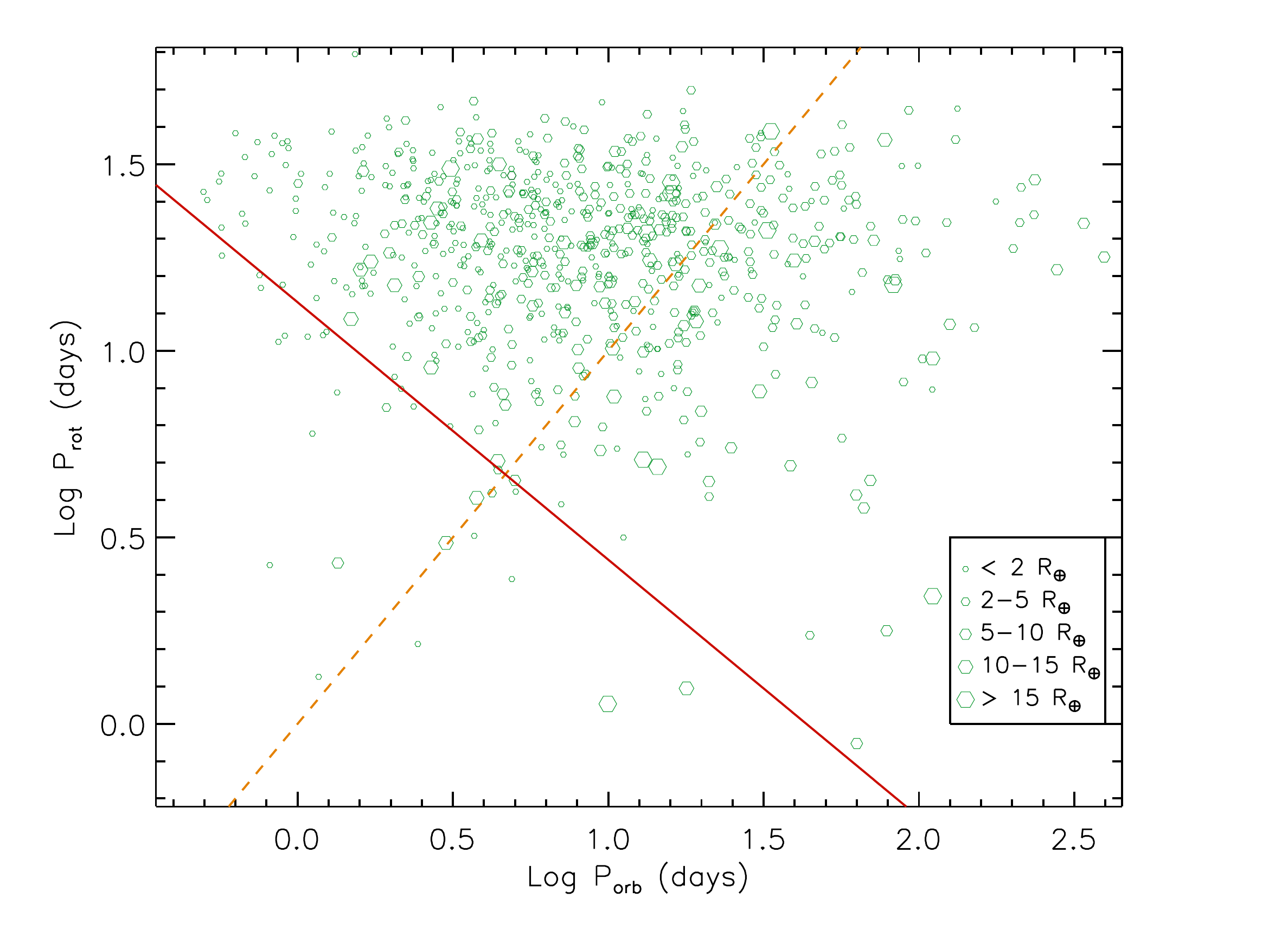} 
\caption{Rotation period vs. the orbital period of the innermost planet for the sample of candidate transiting planets considered by \citet{McQuillanetal13}. The size of the green symbols indicates the radius of the planet as specified in the legend. The red solid line indicates the lower boundary of the distribution as determined by \citet{McQuillanetal13}, while the dashed orange line marks the locus of synchronized stellar rotation ($P_{\rm rot} = P_{\rm orb}$). }
\label{protvsporb}
\end{figure}

A general introduction to tides in stars and planets is provided by, e.g., \citet{Zahn08}, \citet{Mathisetal13}, \citet{Zahn13}, and \citet{Ogilvie14}.
A remarkable observational result that has been connected to the action of tides is the  dearth of close-in planets, i.e. with orbital period $P_{\rm orb} \la 3-5$~d, around rapidly rotating stars, i.e,  with a rotation period $P_{\rm rot} \la 3-5$~d, among Kepler candidates \citep[see Fig.~\ref{protvsporb} and][]{McQuillanetal13,WalkowiczBasri13}. 

\citet{TeitlerKonigl14} proposed an interpretation based on the tidal evolution of the orbits of close-in planets. Their model  included the evolution of the rotation of the host stars  under the action of tides,  stellar wind and internal coupling between the radiative interior and the convection zone. In their model,  the tidal interaction between stars and planets is quite strong with a stellar modified tidal quality factor  $Q^{\prime}_{*} \simeq 10^{5}-10^{6}$, leading  a remarkable fraction of close-in planets to  end their evolution by falling into their hosts. 

\begin{figure}[ht!]
\centering
\includegraphics[width=90mm,angle=0]{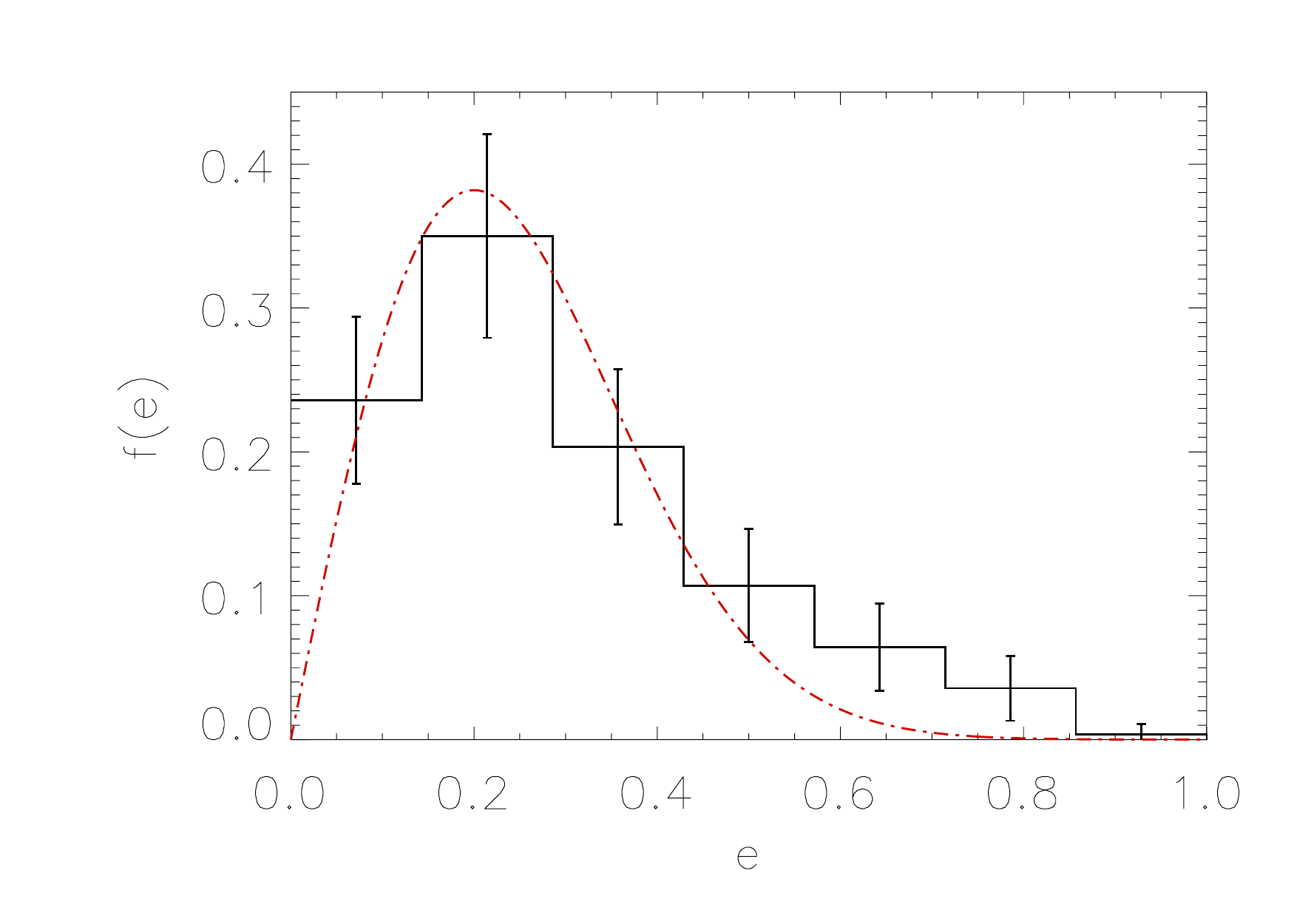} 
\caption{Distribution of the eccentricity of the planets with mass $> 0.1$~M$_{\rm J}$, periastron distance between 0.1 and 10~AU, and $e \geq 0.07$. The histogram is based on the data  extracted on July 21 2014 from the Exoplanets Data Explorer (http://exoplanets.org). The semiamplitudes of the error bars correspond to twice the square root of the number of  planets in each bin. The red dot-dashed line is proportional to a Rayleigh function with $\sigma = 0.2$  \citep[cf. Eq.~(\ref{rayleigh}) in the present paper and Fig.~6 in][]{WuLithwick11}.}
\label{ecc_distr}
\end{figure}

A different explanation was proposed by \citet{LanzaShkolnik14} by considering that most of the Kepler planetary candidates have radii $R_{\rm p} \la 3-4$~R$_{\earth}$ and are probably members of multi-planet systems. In such systems secular chaotic interactions generally develop. They can induce large excursions in the  orbital eccentricity of the innermost planet, while its semimajor axis stays nearly constant \citep{Laskar97,Laskar08,WuLithwick11}. When  those systems reach a stationary dynamical regime, the distribution function of the eccentricity of the innermost planets $f(e)$ is predicted to follow  a Rayleigh distribution, i.e.:
\begin{equation}
f(e) \propto e \exp \left( -\frac{e^{2}}{2\sigma^{2}} \right),
\label{rayleigh}
\end{equation} 
where the standard deviation $\sigma$ depends on the mass of the planets and their perturbers. In Fig.~\ref{ecc_distr} we plot the observed eccentricity distribution for planets with a projected mass $\geq 0.1$~M$_{\rm J}$, i.e., massive enough to allow a sufficiently accurate determination of the eccentricity from their spectroscopic orbits. Only values of $ e > 0.07$ have been considered because lower values may be spurious and result from  errors or noise in  radial velocity measurements. To reduce the effects of tides that circularize the orbits, only planets with a periastron distance between 0.1 and 10~AU have been considered as in \citet{WuLithwick11}. Fig.~\ref{ecc_distr} shows that a Rayleigh distribution function with $\sigma \simeq 0.2$ is not too a bad  approximation to the observed eccentricity distribution. Of course, it  includes systems with very different ages, i.e., for which secular perturbations have acted for different time intervals, so the stationary theoretical distribution is not expected to match it accurately. 
 
When the periastron distance of the innermost planet becomes small enough ($\la 0.05-0.07$~AU), the planet  experiences a strong tidal encounter with its host star and its orbit is shrunk and circularized on a timescale of the order of a few tens of~Myr thanks to tidal dissipation inside the planet itself. Considering a simple model for the probability of such a tidal encounter, \citet{LanzaShkolnik14} can account for the observed distribution of the orbital periods of the planets in the range $3 \la P_{\rm orb} \la 15$~d around main-sequence stars of different ages as estimated from gyrochronology \citep{Barnes07}. In their model,  the probability of finding a planet on a close orbit is higher for older stars because there has been more time for the secular chaos to  excite a higher eccentricity of the planetary orbit before the tidal encounter. Since older stars are slower rotators  owing to  magnetic braking, the probability of finding closer planets is higher around slowly rotating stars. 

A limitation of both K\"onigl \& Teitler's and Lanza \& Shkolnik's models is that the predicted frequency of systems on very close orbits ($P_{\rm orb} \la 1-2$~d) is noticeably different from the observed one. This may indicate that both models are missing some important process that shapes the distribution of such systems, e.g.,  secular perturbations in the case of K\"onigl \& Teitler, or a  final decay of the orbits due to tides inside the stars for Lanza \& Shkolnik. 

The two models require rather different tidal dissipation efficiencies inside the host stars. A value of $Q^{\prime}_{*} \sim 10^{5}-10^{6}$ as considered by \citet{TeitlerKonigl14}, is expected to produce a detectable orbital period variation ($O-C \sim 5-80$~s) in one or two decades in the case of systems with transiting hot Jupiters such as Kepler-17 \citep{BonomoLanza12}, CoRoT-11 \citep{Lanzaetal11}, or WTS-2 \citep{Birkbyetal14}. Therefore, a long-term timing of a sample of transiting hot Jupiters should constrain the efficiency of tidal dissipation in their host stars, indirectly testing the assumptions of Teitler \& K\"onigl's model. It is important to observe a sufficiently large sample of systems to be able to disentangle the tidal orbital evolution from perturbations induced by possible third bodies in the systems themselves. 

In addition to the distribution of the orbital eccentricity of planets with periastron distance $\ga 0.1$~AU, another test of secular chaos may be provided by the distribution of the projected obliquity of the planetary orbits that can be measured through the Rossiter-McLaughlin effect. Thanks to the high precision of the HARPS and HARPS-N spectrographs, this has become feasible also for planets orbiting slowly rotating, old stars with orbital periods of several days \citep[cf.][]{Espositoetal14} for which  secular chaos is expected to induce a significant fraction of highly inclined or even retrograde orbits \citep[e.g.,][]{WuLithwick11}. Considering a sufficiently large sample of such systems, in principle the predictions of the secular chaotic model for the orbital evolution can  be tested\footnote{This possibility has been suggested by Dr. K.~Poppenhaeger in a comment to my talk.}.
\section{Magnetic interactions in stellar chromospheres and coronae}
\label{SPMI}

I shall focus on a few observations and the models proposed for their interpretation. Then I shall briefly consider  the star-planet magnetic interactions from a more general perspective with the help of some recent magnetohydrodynamic (MHD) models.  

\subsection{Some selected observations}
\label{SPMI_obs}

Chromospheric hot spots rotating in phase with the orbit of a hot Jupiter rather than with the stellar rotation have been reported by, e.g., \citet{Shkolniketal03,Shkolniketal05,Shkolniketal08}, and \citet{Gurdemiretal12}. The power emitted by those spots can reach $\sim 10^{20}-10^{21}$~W, i.e., a few percent of the entire chromospheric flux and they are not stationary phenomena because they are observed only in $30-50$ percent of the seasons in the case of HD~179949 and $\upsilon$~And, the systems with the best evidence to date. The spots are not at the subplanetary longitude, but are shifted up to $\sim 170^{\circ}$ in the case of $\upsilon$~And. Several studies have questioned the reality of the phenomenon, possibly owing to its non-stationary nature \citep[cf.][]{Poppenhaegeretal11,Milleretal12,Scandariatoetal13}. Nevertheless, the consideration of those hot spots has greatly stimulated the development of theoretical models for the interaction between stellar and planetary magnetic fields (see Sects.~\ref{analytic_models} and~\ref{numerical_models}).

Another intriguing observation has been reported by \citet{Fossatietal13} who found a very low level of chromospheric emission in WASP-12, WASP-18, and a few other systems with transiting hot Jupiters as observed in the Ca~II H\&K resonance lines. Not only the Ca~II H\&K lines, but also the Mg~II h\&k lines of WASP-12  show no detectable emission in their cores \citep{Haswelletal12}. The explanation proposed by \citet{Haswelletal12} and \citet{Fossatietal13}  is that the lack of flux in the cores of the chromospheric resonance lines is due to absorption by circumstellar material probably evaporated from the very close hot Jupiter having  $P_{\rm orb} \sim 1.1$~d in WASP-12. However, the possibility that the star has a very low intrinsic level of chromospheric activity, even lower than the basal level generally observed in stars of the same spectral type without hot Jupiters, cannot be completely excluded in view of  the lack of detection of photospheric magnetic fields \citep{Fossatietal10a}. Similarly, \citet{Pillitterietal14a} found a very low level of coronal X-ray emission in WASP-18 that is unlikely to be due to circumstellar absorption and suggests a very low  magnetic heating of the outer stellar atmosphere. 

The observations of the transit of the hot Jupiter  WASP-12 with HST in the near ultraviolet ($\approx 254-280$~nm) show an earlier ingress and a deeper transit than in the optical passband, suggesting absorption by some unevenly distributed circumplanetary gas, with a density enhancement preceding the planet along its orbit \citep{Fossatietal10b}. 

Observations of HD~189733, a K-type main-sequence star with a transiting hot Jupiter on a $P_{\rm orb}  \sim 2.2$~d orbit, have detected coronal flares occurring immediately after the egress of the planet from the stellar occultation, i.e. in the phase range $0.55-0.65$, in three different seasons \citep{Pillitterietal11,Pillitterietal14b}. Although the statistics is still limited, such observations suggest a  possible connection between the frequency of stellar flares and the orbital phase of the planet, something possibly reminiscent of the chromospheric hot spot phenomenon, although much more transient in nature \citep[cf. ][ in particular their Sect.~3.2]{Shkolniketal08}. 

In view of the above observations, the main questions to be addressed by theoretical models can be summarized as follows:
\begin{enumerate}
\item[a)] what is the physical process responsible for the energy dissipated in the chromospheric hot spots ?
\item[b)] why are they shifted with respect to the subplanetary longitude ?
\item[c)] what is producing the low level of chromospheric emission in some stars with transiting hot Jupiters ? 
\item[d)] What is producing the asymmetric transit profile in the UV in the case of WASP-12 ?
\item[e)] Is there any mechanism to account for preferential orbital phases for flaring activity as suggested by the observations of HD~189733 ?
\end{enumerate}

\subsection{Analytic  models of chromospheric and coronal interactions}
\label{analytic_models}

The first, seminal work on the possibility of star-planet interaction was that of \citet{Cuntzetal00}. Since that time, more detailed investigations have been performed, also pointing out similarities and differences with the Sun-planet interactions.  
In the Solar system, planets are located in a region where the velocity of the solar wind $v_{\rm w}$ is greater than the local Alfven velocity $v_{\rm A} = B/\sqrt{\mu \rho}$, where $B$ is the magnetic field in the interplanetary space, $\mu$ the permeability of the plasma, and $\rho$ its density. In this regime, a bow shock forms at the magnetospheric boundary in the case of a planet having an intrinsic magnetic field, as for the Earth. On the other hand, most close-in planets are probably located in a region where the velocity of the stellar wind is smaller than the local Alfven velocity \citep{Preusseetal05}, leading to a rather different regime.

The first models proposed to account for chromospheric hot spots were based on the analogy with the Jupiter-Io system. Io orbits inside the magnetosphere of Jupiter and is a continuous source of Alfven waves that propagate inward and reach Jupiter's atmosphere where they dissipate energy producing a hot spot that rotates in phase with the orbital motion of the satellite. Io is very likely not possessing an intrinsic magnetic field, but it has an electrically conducting body. In this regime, the excitation and propagation of Alfven waves along the magnetic field lines of Jupiter's magnetosphere is described by the so-called unipolar inductor model \citep[for a review see][]{Sauretal13}. 

The same model has been applied to account for the stellar chromospheric hot spots and is capable of explaining the large phase lag between a planet and the associated hot spot thanks to the  travel time needed for the waves to propagate from the planet to the star \citep{Preusseetal06,Koppetal11}. The location of the planet inside the sub-alfvenic wind region is a fundamental requisite for  this model because in the super-alfvenic domain any Alfven wave excited close to the planet is blown away by the wind moving faster than its phase velocity and cannot propagate down to the star.  \citet{Sauretal13}  find that about 30 percent of  hot Jupiters can excite Alfven waves that may reach their host stars with an  available power  up to $10^{18}-10^{19}$~W.  Those estimates are based on the typically measured stellar fields as derived from spectropolarimetry \citep[e.g. ][]{Moutouetal07,Faresetal12,Vidottoetal14} and adopted planetary fields of the order of that of Jupiter ($\approx 10-20$~G), and are  insufficient by $2-3$ orders of magnitude to account for the emission of the hot spots as observed in, e.g., HD~179949. 

A different approach has been introduced  by \citet{Lanza08,Lanza09,Lanza12,Lanza13}. It assumes that in the domain from  the outer stellar atmosphere up to the distance of the planet: a)  gravity and  plasma pressure are negligible in comparison to the Lorentz force; and b) the velocities of the plasma and of the planet are much smaller than the Alfven velocity. Therefore, the system can be assumed to be in a stationary, magnetohydrostatic equilibrium. In this regime, the equilibrium is everywhere characterized by the force-free condition, i.e., the Lorentz force ${\bf \cal L }= {\bf J} \times {\bf B} = 0$, where ${\bf J } = \mu^{-1} (\nabla \times {\bf B})$ is the current density. In other words,  $(\nabla \times {\bf B}) \times {\bf B} = 0$. This means that the curl of the field is everywhere parallel to the field itself and we can write $\nabla \times {\bf B} = \alpha {\bf B}$, where $\alpha$ is a scalar that is constant along each field line as immediately  follows by taking the divergence of both sides of the previous equation and considering that ${\bf B}$ is solenoidal: ${\bf B} \cdot \nabla \alpha = 0$. 

\begin{figure}
\centerline{
\includegraphics[width=60mm,height=75mm]{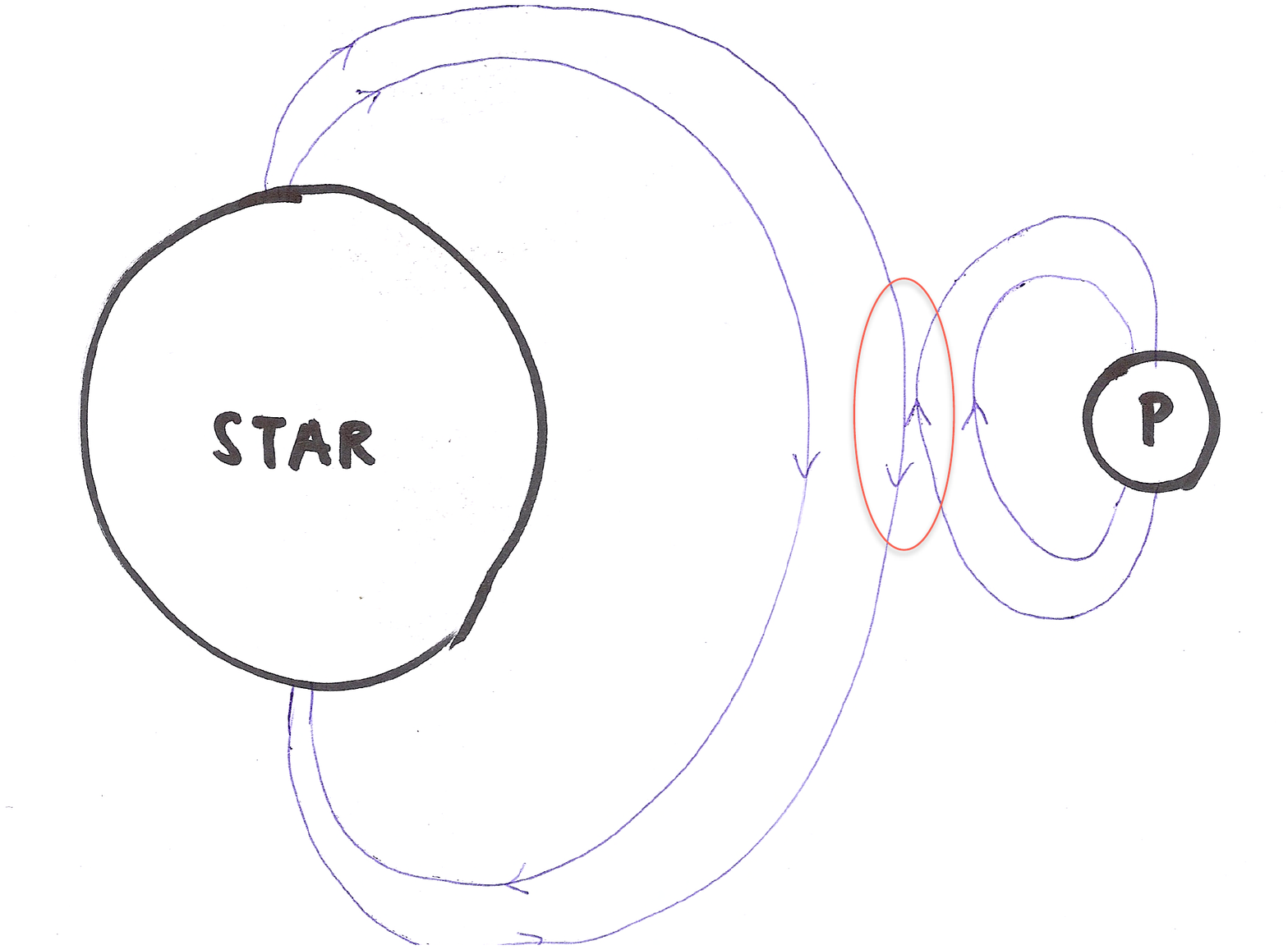} 
\includegraphics[width=60mm,height=83mm]{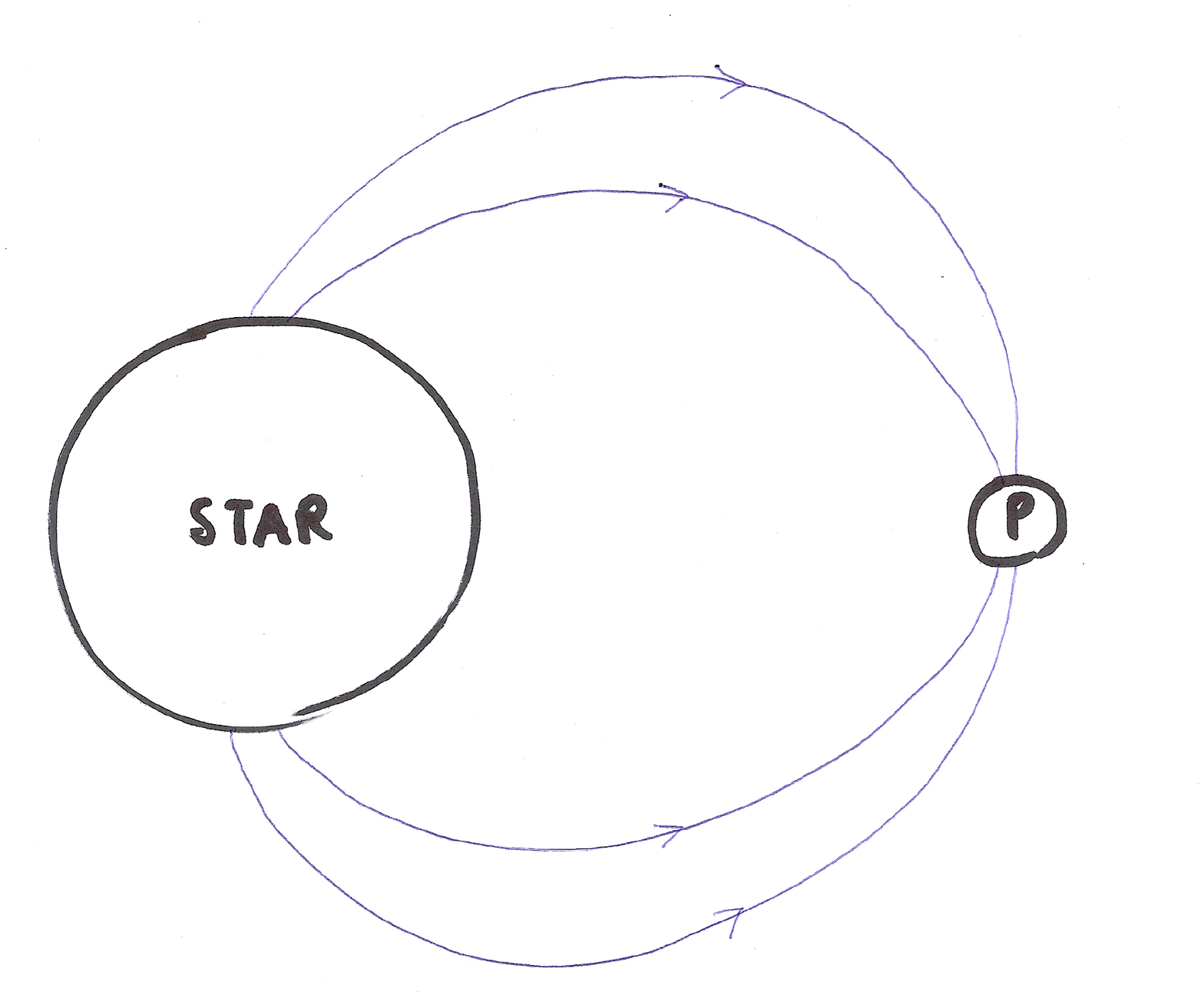}} 
\caption{Left: Magnetic topology consisting of two separated flux line systems for the stellar coronal field and the planetary field. The red oval marks the domain where oppositely directed magnetic field lines of the two systems reconnect. Right: Magnetic topology with a loop interconnecting the star and the planet (P).}
\label{magnetic_configs}
\end{figure}

In general, two different magnetic topologies are possible, if the planet has an intrinsic magnetic field (cf.~Fig.~\ref{magnetic_configs}). 
When the boundary conditions at the base of the stellar corona are such that $\alpha \not= 0$ along the field lines that reach the planet, the coronal field and the planetary field are topologically separated because the planetary field is potential, so that $\alpha=0$ along its field lines (cf. Fig.~\ref{magnetic_configs}, left panel). In other words, if there were a field line interconnecting the star and the planet, $\alpha$ would be constant along it, but this is not possible given that $\alpha \not= 0$ on the star, while $\alpha = 0$ on the surface of the planet because currents are not allowed to propagate through its mostly neutral atmosphere. Therefore, the stellar and the planetary fields are  allowed to interact only at the boundary of the planetary magnetosphere where magnetic reconnection takes place if the field lines have different directions. 

The radius of the planetary magnetosphere $R_{\rm m}$  is determined by magnetic pressure equilibrium between the stellar and planetary fields  because the ram pressure of the stellar wind is assumed to be negligible \citep{Lanza09}. In this regime, the power dissipated by the reconnection process is: 
\begin{equation}
P_{\rm d} \simeq \gamma \frac{\pi}{\mu} \left[ B_{*} \left( a, \frac{\pi}{2} \right) \right] ^{2} R_{\rm m}^{2} v_{\rm rel},
\label{reco_power}
\end{equation}
where $0 < \gamma < 1$ is an efficiency parameter depending on the relative orientation of the interacting field lines,  $B_{*} \left( a, \frac{\pi}{2} \right)$  the coronal field at the distance $a$  of the planet's orbit assumed to be circular and on the equatorial plane of the star (colatitude $\theta = \frac{\pi}{2}$),  and $v_{\rm rel}$  the relative velocity between the coronal and the planetary field lines that corresponds to the orbital velocity of the planet for a slowly rotating star. Considering typically measured fields in active planetary hosts, $B_{*} (R_{*})= 10-40$~G at the surface of the star ($r=R_{*}$), and a planetary field at its poles $B_{\rm p} = 10$~G, we obtain for a typical hot Jupiter around a sun-like star with $v_{\rm rel} =100-200$~km/s, $R_{\rm m} \sim 4-5$~$R_{\rm p}$ and $P_{\rm d} \sim 10^{17}-10^{19}$~W, where $R_{\rm p}$ is the radius of the planet \citep[cf. ][]{Lanza12}\footnote{The given power range accounts for different stellar field configurations with the lower values associated with a potential field and the higher ones with non-linear force-free fields.}.    This is insufficient by $2-3$ orders of magnitudes to account for the power emitted by  hot spots, although a twisted stellar force-free field can easily account for the observed phase lag between the spot and the planet \citep[cf.][]{Lanza08,Lanza09}. 

\citet{Lanza12} proposed that the perturbation of the stellar coronal field induced by the motion of the planet may trigger a release of the energy previously stored in the coronal field itself by reducing the magnetic helicity of its configuration. Previous order-of-magnitude calculations by \citet{Lanza09} showed that the reduction of magnetic helicity associated with the motion of the planet is a promising mechanism to account for the power released in hot spots. Moreover, a periodic modulation of the magnetic helicity is possible in a long-lived magnetic loop that reaches to the orbit of the planet. This may trigger flares with the characteristic periodicity of the orbital period as suggested by \citet{Pillitterietal14b} in the case of HD~189733. 
The conjectured mechanism is analogous to that proposed for solar flares where the minimum-energy configuration, corresponding to a potential magnetic field for the given photospheric boundary conditions, can be reached only when the magnetic helicity of the coronal field is substantially reduced as a result of the emergence of new magnetic flux, motions at the boundary, or a loss of magnetic flux associated with a coronal mass ejection. 

\begin{figure}[ht!]
\centering
\includegraphics[width=90mm,angle=0]{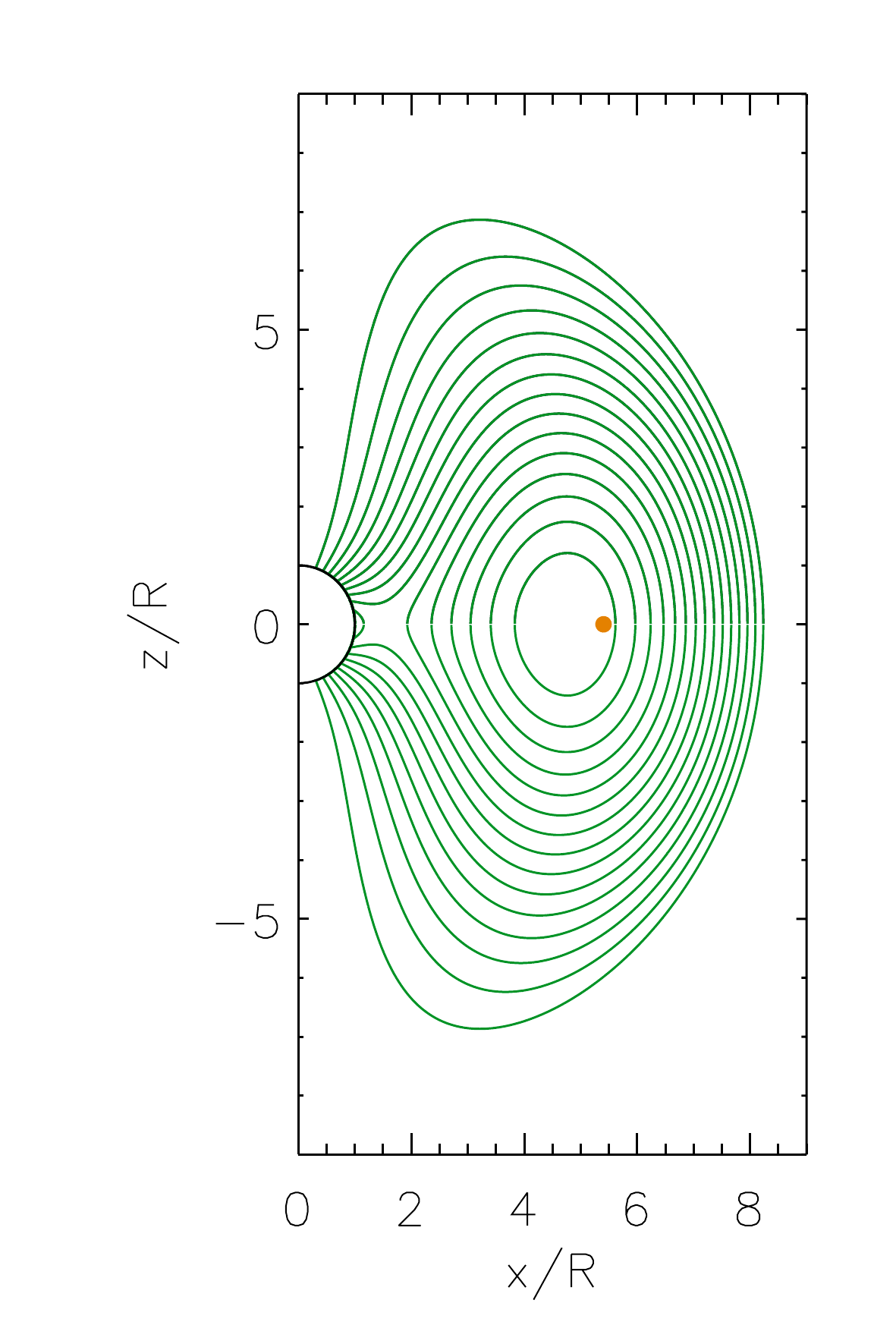} 
\caption{Meridional section of an axisymmetric coronal force-free field with an azimuthal magnetic flux rope consisting of closed field lines that encircle the star. The projections of the field lines onto the meridional plane are shown as green solid lines, while the projection of the stellar surface is given by the black semicircle centered at the origin of the reference frame.  The unit of measure is the radius $R$ of the star with the Cartesian coordinates $x$ and $z$ measured from the origin at the center of the star in the equatorial plane and along the spin axis, respectively. The filled orange circle marks the position of a planet orbiting in the equatorial plane inside the azimuthal flux rope. }
\label{hd179949_linear_model}
\end{figure}

The reduction of the magnetic helicity associated with the orbital motion of the planet will in general drive the field toward a linear force-free configuration, that is a field with a constant $\alpha$  along all the different field lines \citep[see ][]{Lanza09,Lanza12}. Some linear configurations are particularly interesting because they have a rope of azimuthal flux that encircles the star with closed field lines that do not intersect the surface of the star (see Fig.~\ref{hd179949_linear_model})\footnote{Also non-linear axisymmetric force-free configurations can have an azimuthal flux rope, but for simplicity here we shall limit ourselves to the linear case that represents the minimum energy configuration for a given total helicity.}. If the planet is inside the flux rope, the energy released by magnetic reconnection cannot reach the star and no hot spot is produced. Nevertheless, the intense stellar ultraviolet radiation and the energy made available by magnetic reconnection can power the evaporation of the planetary atmosphere \citep{Lanza13} that fills the flux rope with a hot plasma that subsequently cools and condenses into clumps with a typical temperature of the order of $10^{4}$~K. They may effectively absorb in the core of the Mg II h\&k or Ca II H\&K lines, accounting for the extremely low flux level in the cores of those lines in the case of some systems with close-in massive planets (see Sect.~\ref{SPMI_obs}). The formation of an azimuthal flux rope depends on the boundary conditions and the value of the helicity of the stellar field in the stationary configuration. This depends, in turn, on the processes that change magnetic helicity and the dissipation  induced by the planet \citep[see][for details]{Lanza09}. 

On the other hand, when the magnetic helicity of the stellar field is low, the field is close to a potential configuration and there can be field lines interconnecting the base of the stellar corona  with the surface of the planet (cf. Fig.~\ref{magnetic_configs}, right panel). These interconnecting loops are continuously stressed by the orbital motion of the planet, storing energy in excess of the minimum energy corresponding to the potential field. If such an energy can be dissipated at the same rate it is stored into the loop, the field can attain a stationary configuration and the interconnecting loop becomes a persistent feature of the system. The dissipated power can be easily calculated from the flux of the Poynting vector across the base of the interconnecting loop at the pole of the planet \citep{Lanza13} and is:
\begin{equation}
P_{\rm int}  \simeq \frac{2\pi}{\mu} f_{\rm AP} R_{\rm p}^{2} B_{\rm p}^{2} v_{\rm rel},
\end{equation}
where $f_{\rm AP}$ is the fraction of the planetary surface crossed by the interconnecting field lines \citep[typically $f_{\rm AP} \sim 0.1-0.2$, cf.][]{Adams11}. The dissipated power reaches $10^{20}-10^{21}$~W for the typical  parameters adopted above, easily accounting for the emission  of hot spots. The probability of having a stellar axisymmetric potential field  is higher when the magnetic  helicity is low, that is in the phases of minimum of the stellar cycle when the large-scale field is mainly poloidal according to  solar-like dynamo models\footnote{Note that a non-axisymmetric field would produce a chromospheric hot spot with a periodicity different from the orbital period of the planet \citep[see][for details]{Lanza12}.}. This could explain why chromospheric hot spots are not stationary features since they can be preferentially produced during the minima of a stellar activity cycle. Note that in the case of F-type stars magnetic cycles can be as short as a couple of  years as observed in the case of $\tau$~Bootis \citep{Faresetal09}.

Finally, it is interesting to mention models proposed to explain the asymmetric and deeper transits of WASP-12 in the ultraviolet. Since the planet is close to filling its Roche lobe, \citet{Laietal10} proposed that a stream of matter flowing toward the star through the inner Lagrangian point and tilted in the direction of the orbital motion by the conservation of the angular momentum could absorb light  in the UV just before the transit when its column density is the highest. Alternatively, a bow shock may develop in front of the planet due to its orbital motion inside the stellar corona or wind. This idea has been further explored by \citet{Vidottoetal10}, \citet{Llamaetal11}, and \citet{Vidottoetal11} and can in principle provide an estimate of the strength of the planetary field, if the stellar coronal field intensity is known by extrapolating the field observed on the photosphere through spectropolarimetric techniques.

\subsection{Numerical models}
\label{numerical_models}

Several results of analytic models have been confirmed by numerical MHD models. They can include the effects of finite plasma pressure and gravity, and simulate non-stationary configurations going beyond the very simple assumptions of analytic models. On the other hand, numerical models cannot fully resolve all the physically relevant lengthscales of the systems, therefore they include simplified (sometimes even crude) parameterization of the processes over sub-grid scales. Nevertheless, the  recent progress in the field of stellar MHD simulations \citep[e.g., ][]{Brunetal13} makes them truly promising to explore star-planet magnetic interactions as well \citep[e.g.,][]{Strugareketal12,Strugareketal14}. 

Interesting results were  obtained by \citet{Cohenetal09,Cohenetal11a,Cohenetal11b} who confirmed the possibility of forming interconnecting magnetic loops that can be stressed by the orbital motion of the planet and suggested that  enough energy is available to account for the chromospheric hot spots in the case of HD~179949 or $\upsilon$~And. Those simulations  included  a realistic large-scale field geometry as obtained by extrapolating the field measured on the stellar photosphere through spectropolarimetric techniques. They were not limited to simple, stationary configurations and were extended to treat time-dependent events such as  the interaction of a coronal mass ejection with a planet. 

Recent work focussed on M-type dwarf stars whose habitable zones are closer than in the case of the brighter sun-like stars. By extrapolating spectropolarimetric field observations, it has become possible to study the large-scale geometry of their Alfven surfaces, i.e.,  the transition surfaces separating the sub-alfvenic wind regions from the super-alfvenic ones \citep[e.g., ][]{Vidottoetal14b}. \citet{Cohenetal14} showed that the formation of a bow shock in the super-alfvenic wind regime helps to protect the planet from the effect of mass loss, even when the evaporation induced by stellar irradiation is strong, because most of the planetary field lines along which the evaporating plasma flows are bent and closed back in the magnetospheric tail. On the other hand, in the sub-alfvenic regime, a substantial fraction of the field lines are open and connected to the wind field lines, thus  the evaporating plasma is free to escape from the planet producing a significantly larger mass loss. Since the Alfven surface, as reconstructed by photospheric field extrapolation, is generally irregular and far from a spherically symmetric shape,  the MHD regime can switch from sub- to super-alfvenic and viceversa several times along the orbit of the planet. Adding the time variability of the stellar field (long-term activity evolution and/or short-term stellar activity cycles) this makes a planet experience rather different regimes during its lifetime. 

\section{Evaporation of planetary atmospheres}
\label{evaporation_sec}

\citet{VidalMadjaretal03,VidalMadjaretal04,VidalMadjaretal08}, \citet{Linskyetal10}, and others found evidence of a deeper transit in the Lyman-$\alpha$ and other far-ultraviolet lines (e.g., of SiIII~$\lambda 120.65$~nm, CII~$\lambda 133.45$ and~$\lambda 133.57$~nm) with depths $2-3$ times greater than in the optical bandpass for HD~209458, HD~189733, and WASP-12. A remarkable time variability has been reported  by \citet{Lecavelierdesetangsetal12} in the case of HD~189733. The absorbing material extends beyond the planetary Roche lobe and has velocities ranging from several tens to $\approx 100-150$ km/s. These observations are  interpreted as evidence of evaporation of planetary atmospheres. \citet{Haswelletal12} extended the approach to the near-ultraviolet lines finding evidence of evaporation  in  WASP-12b. 

The stellar extreme-ultraviolet (EUV) flux ($\lambda \la 91$~nm) has been identified as the main source of energy to power the evaporation. It depends on the spectral type and the rotation rate of the star that determine the overall heating level of the transition region and the corona emitting in that  passband. The flux can vary remarkably along a stellar activity cycle (at least by a factor of $2-3$ in the Sun) and be enhanced during flares. Owing to absorption by interstellar Hydrogen, it is difficult to measure the EUV flux, so different estimates of it have been proposed for stars of various ages and spectral types \citep[e.g.,][]{Ribasetal05,Lecavelierdesetangs07,Sanz-Forcadaetal11,Linskyetal13}.

\begin{figure}[h!]
\centering
\includegraphics[height=120mm,width=90mm,angle=0]{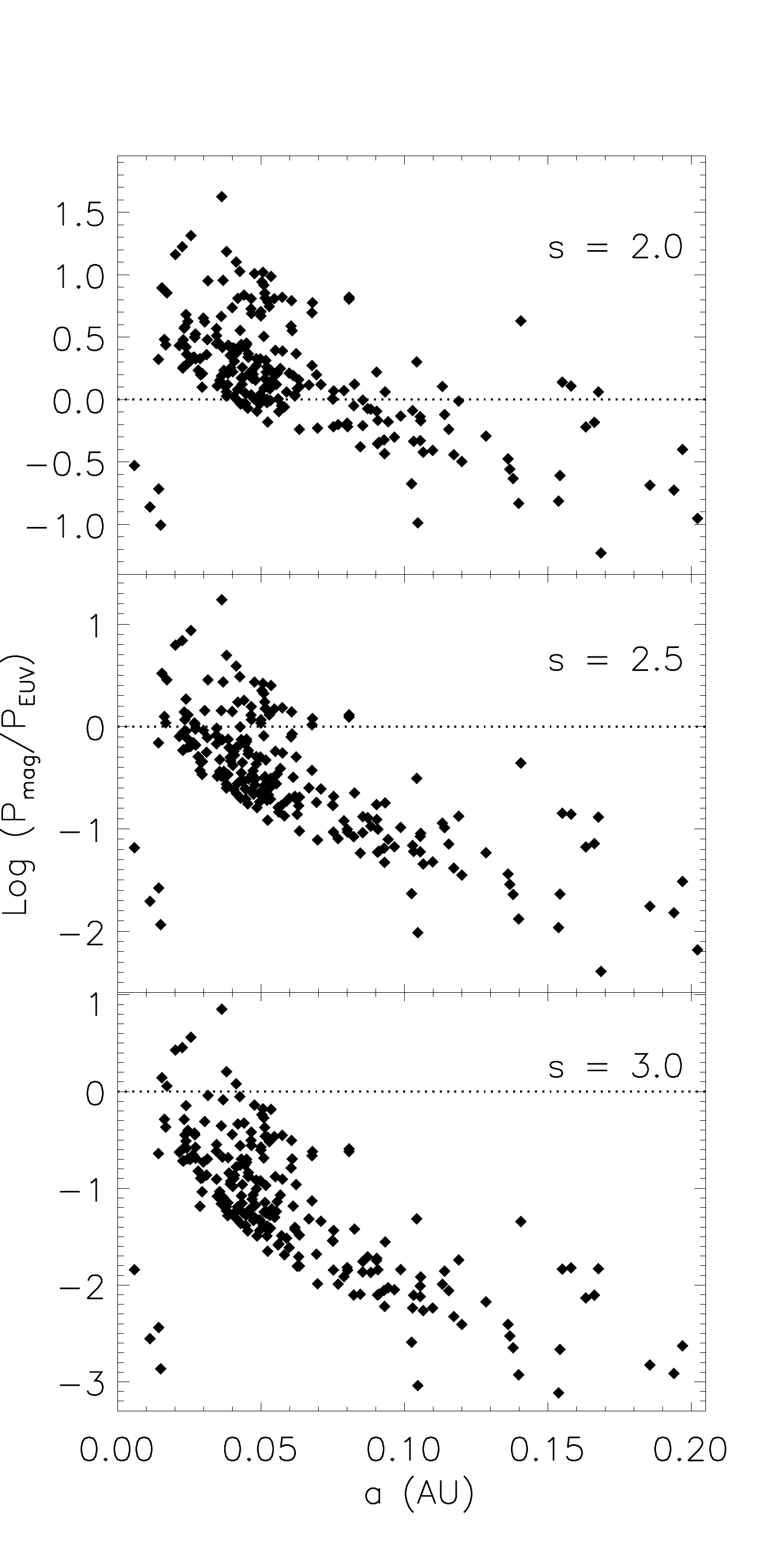} 
\caption{Upper panel: ratio of the power delivered by magnetic reconnection into the planetary atmosphere to the power of the stellar EUV irradiation vs. the orbital semimajor axis. The parameter  $s=2.0$ corresponds to a stellar magnetic field having radially directed field lines (see the text). The horizontal dotted line marks  $P_{\rm mag} = P_{\rm EUV}$, i.e., the two sources of evaporation having the same power. Middle panel: the same as the upper panel, but for a force-free field with $s=2.5$. Lower panel: the same as the upper panel, but for a potential coronal field ($s=3.0$). }
\label{evaporation}
\end{figure}

Recently, \citet{Buzasi13} and \citet{Lanza13} proposed an additional source  to power evaporation, i.e., the energy released by magnetic reconnection between stellar and planetary fields at the boundary of the planetary magnetosphere (cf. Sect.~\ref{analytic_models} and Eq.~\ref{reco_power}). Reconnection accelerates electrons that are conveyed down along the magnetic field lines reaching the base of the planetary exosphere, where they  induce evaporation and chemical reactions. An estimate of the minimum available power in units of that of the EUV flux \citep{Lecavelierdesetangs07} is shown in Fig.~\ref{evaporation} as a function of the orbital semimajor axis for different configurations of the stellar magnetic field. Specifically, we assume that the coronal field depends on the radial distance $r$ from the center of the star as $B_{*}(r) = B_{*}(R_{*}) (r/R_{*})^{-s}$, where $B_{*}(R_{*})$ is the field at the surface of the star ($r=R_{*}$) and $ 2 \leq s \leq 3$ is a parameter depending on the field geometry, ranging from $s=2$ for a purely radial field to $s=3$ for a potential field; intermediate values are associated with force-free fields \citep[cf. ][]{Lanza12,Lanza13}. Assuming a planetary field comparable with that of Jupiter ($B_{\rm p} =10$~G), a stellar field $B_{*}(R_{*})= 10$~G, as indicated by spectropolarimetric observations \citep{Moutouetal07,Faresetal12}, and $\gamma =0.5$, we find that planets closer than $0.03-0.05$~AU experience a  magnetic heating greater than  the estimated  stellar EUV flux. The corresponding mass loss is therefore significantly increased, although the effect  is  generally not large enough to completely ablate the closest planets during the main-sequence lifetime of their host stars \citep[see][]{Lanza13}. 

Given the remarkable time variability of  stellar magnetic fields, we expect that magnetically powered evaporation is not constant. \citet{Kawaharaetal13} suggest that the evaporation rate in the very close transiting object KIC~12557548 ($P_{\rm orb} \sim 0.6536$~d) be modulated by the rotation of its spotted star. This could  arise from the rotational modulation of the EUV flux coming from its active regions, but also a magnetically-induced evaporation may be invoked because the stellar magnetic field at the boundary of the planetary magnetosphere is modulated as well. 

Energetic electrons accelerated by reconnection and conveyed into the planetary exosphere can produce chemical reactions that lead to the formation of molecules such as C$_{2}$H$_{2}$, C$_{2}$H$_{4}$, NH$_{3}$, C$_{6}$H$_{6}$. They have potential astrobiological relevance and may  not be produced with the same efficiency by EUV radiation because photons have no electric charge  \citep[cf.][]{Rimmeretal14a,Rimmeretal14b,Starketal14}.

\section{Star-planet interaction and evolution of stellar rotation}
\label{starrotation}

\citet{Pont09}, \citet{Brownetal11}, and \citet{Bolmontetal12}, among others, proposed that tides can spin up planet-hosting stars, especially in the case of giant planets orbiting slowly rotating late-type stars. A quantitative study, however, is made difficult by our limited knowledge of the tidal coupling in star-planet systems  and its dependence on the stellar rotation rate \citep[see Sect.~\ref{tides} and][]{OgilvieLin07}. 
Moreover, some theoretical models predict that the braking of stellar rotation in late-type stars produced by their magnetized winds can be reduced thanks to the interaction with a close-in planet \citep[e.g.,][]{Cohenetal10,Lanza10}.  

\citet{Brown14} compared the ages estimated for stars with close-in planets by means of gyrochronology and isochrone fitting, but the large intrinsic errors prevented him from reaching definite conclusions. Some progress should be possible by applying  asteroseismology to improve the determination of stellar ages \citep[cf.][]{Angus14}, independently of the rotation rate. 

Another approach exploits the presence of a distant, presumably co-eval, visual companion whose age can be estimated from its X-ray coronal flux. \citet{PoppenhaegerWolk14} summarize the results obtained so far and find that CoRoT-2 and, to a lesser extent, HD~189733, appear to rotate faster than expected from gyrochronology on the basis of the estimated ages of their visual companions, respectively. 

\section{Evidence of star-planet interaction in photospheric activity ?}
\label{starspots}

The possibility that photospheric active regions be associated with close-in planets has also been considered. \citet{Santosetal03} reported the intriguing case of HD~192263, a K2V star accompanied by a planet with an orbital period of 24.35~d and a projected mass of 0.72~M$_{\rm J}$, that showed during some seasons a photometric modulation with the same period of its planet. This led \citet{Henryetal02} to question the  planetary nature of the radial velocity modulation, but its long-term stability supports the reality of the planet. The coincidence $P_{\rm rot} \simeq P_{\rm orb}$ would be easily understood in terms of tidal synchronization of the stellar rotation with the orbital motion, but the long orbital period makes the tidal interaction remarkably weak in this  system. Of course, the possibility that the stellar rotation period is close to the orbital period of the planet by chance cannot  be excluded, but an alternative exotic interpretation is that the formation and evolution of starspots  are in some way coupled with the orbital motion of the planet. Further similar cases have been reported. 

\citet{Walkeretal08} studied the case of $\tau$~Bootis, an F-type star rotating synchronously with the orbital period of its massive planet ($P_{\rm rot} \sim P_{\rm orb} \sim 3.3$~d). It showed a long-lived active region that has maintained an approximately constant phase lag of $\approx 75^{\circ}$ with respect to the planet for at least $\approx 5$~years. 

\citet{Lanzaetal09b} found that in the F-type dwarf star CoRoT-4 the rotation of the photospheric starspots was approximately synchronous with the orbital period of the planet ($P_{\rm orb} \sim 9.2$~d) and there was a spot close to the subplanetary longitude during most of the CoRoT observations covering a time span of $\sim 60$~d. Tides are weak in the CoRoT-4 system because of the long orbital period and the relatively low mass of the planet ($\sim 0.75$~M$_{\rm J}$). 

Another interesting case is that of CoRoT-6, a F9V star  that has a planet with a mass of $\sim 3$~M$_{\rm J}$ and $P_{\rm orb} = 8.89$~d, while the mean stellar rotation period is $P_{\rm rot} \sim 6.35$~d. \citet{Lanzaetal11b} found that starspot activity was enhanced in the active regions when they crossed a longitude that was trailing the subplanetary longitude by $\sim 200^{\circ}$ during the $\sim 150$~d of CoRoT observations. 

A phenomenon reminiscent of such a behaviour was found by \citet{Bekyetal14} in the transiting systems HAT-P-11 and Kepler-17. The rotation periods of those stars are not synchronized with the orbital periods of their  planets, respectively, but are close to an integer multiple of them. Precisely, $P_{\rm rot} \sim 6 P_{\rm orb}$ in the case of HAT-P-11 and $P_{\rm rot} \sim 8 P_{\rm orb}$ in Kepler-17. The intriguing phenomenon is that, in spite of the imperfect commensurability of  the mean rotation and orbital periods, there are starspots that rotate with an almost perfect commensurable period in both systems. This is demonstrated  by the recurrence of their occultations by the planets at  the same orbital phases  along  datasets spanning more than $\approx 800$~days. The observed recurrence rates of those starspots are about 2.3 times the values expected in the case of a random distribution of their rotation rates. 

Although a global estimate of the probability that the coincidences reported above occurred by chance has not been given yet, the cases based on CoRoT and especially Kepler long-term observations suggest that a close-in massive planet may in some way modulate the starspot activity in the photosphere of its host star. Other intriguing observations refer to preferential orbital phases for an enhancement of the stellar flux variability \citep{Paganoetal09}  and a short-term starspot cycle in CoRoT-2 with a period close to ten times the synodic period of the planet with respect to the mean stellar rotation period \citep{Lanzaetal09a}. A similar short-term spot cycle was observed in Kepler-17, but no commensurability with the synodic period was found in that case \citep{BonomoLanza12}. 

From the theoretical side, it is difficult to find a  physical mechanism to account for a starspot activity phased to the orbital motion of a close-in planet. In the case of close active binaries, the strong tides in the almost synchronized component stars might interact with large scale convection and lead to active longitudes that rotate almost in phase with the orbital motion \citep[cf.][ for the role of large-scale convection in the formation of active longitudes]{Weberetal13}. Magnetic flux tubes close to the base of the convection zone could also be de-stabilized by the tidal deformation of the stars producing active longitudes \citep{HolzwarthSchussler03a,HolzwarthSchussler03b}. 
However, it is unclear whether those mechanisms may account for the phenomena observed in planetary systems because the mass of a hot Jupiter is of the order of $10^{-3}$ of the mass of a stellar companion, so  tidal effects are dominated by stochastic convective motions inside the star. However, numerical simulations of the dynamo action in systems affected by tides are beginning to be performed and they have the potential to shed light  on those effects  \citep[cf. ][]{CebronHollerbach14}. 

It has also been conjectured  that the  interaction between the planetary and the stellar magnetic fields may modulate the total magnetic helicity of the stellar field that, in turn, affects the operation of the stellar dynamo \citep[cf. ][]{BrandenburgSubramanian05}. Helicity  is mainly dissipated by eruptive processes in the corona such as  mass ejections and large flares, so it is conceivable  that a planet moving close to its star may modulate the occurrence of those phenomena and indirectly affect the stellar dynamo \citep[see ][ for details]{Lanza08,Lanza09,Lanza11}.

\section{Conclusions}
\label{conclusions}

Star-planet interactions are a fascinating field. I have reviewed some topics that have mostly impressed me and that are promising for further investigations both observationally and theoretically. Planetary systems offer the  opportunity to test tidal theories in previously unexplored regimes with extreme mass ratio between the two bodies, eccentric orbits, and stellar rotation far away from synchronization. Tides in multi-body systems can also be investigated. I have discussed in some detail  one application of the tidal models to account for the intriguing distribution of Kepler candidate systems in the $P_{\rm rot}-P_{\rm orb}$ plane. 

Magnetic interactions in systems with close-in planets can take place in a regime characterized by a local Alfven speed smaller than the plasma (stellar wind) speed. This allows a perturbation excited by the planet to propagate down to its host star in a regime not observed in our Solar system. Simple MHD models predict different magnetic configurations in star-planet systems. Configurations characterized by an azimuthal magnetic flux rope may store the plasma evaporating from a close-in  planet possibly accounting for the presence of circumstellar absorption as suggested by the observations of some systems. On the other hand, magnetic loops interconnecting the star with the planet may be stressed by the orbital motion of the planet and dissipate powers up to $10^{20}-10^{21}$~W that may account for chromospheric hot spots moving in phase with the planet rather than with stellar rotation. The energy dissipated by magnetic reconnection can  play a relevant role in the evaporation of planetary atmospheres, especially for close-in ($ \la 0.05$~AU) planets.

Photospheric spots and coronal flares associated with a close-in planet have  been suggested by recent observations. Moreover, the magnetic and tidal interactions between a planet and its host may affect the angular momentum evolution and the activity level of the star making standard gyrochronology difficult to apply. 

New observations and theoretical models are mandatory for making progress in this field. A long-term monitoring of an extended sample of targets is necessary to confirm the phenomena suggested by the  observations discussed above. The most severe limitation in  the case of CoRoT and Kepler is the scarcity of bright stars with close-in massive planets, the only systems for which these subtle effects can be detected. The recently selected space mission PLATO is therefore particularly interesting because it is dedicated to the detection, characterization, and long-term (at least $2-3$ yr) monitoring of planetary systems orbiting bright stars \citep{Raueretal14}. For those stars, asteroseismology can allow us to determine the age with typical errors of $10-15$ percent, thus allowing to test and calibrate gyrochronology for planetary hosts (cf. Sect.~\ref{starrotation}).


%
%



%
%




%
%


\acknowledgments{
I wish to thank the Scientific Organizing Committee of Cool~Stars~18 for their kind invitation to review star-planet interactions and for the organization of such a stimulating meeting. I am  grateful to Drs.~Airapetian and Shkolnik for interesting discussions. 
}

\end{document}